\documentclass[english,twoside,a4paper,10pt]{article}

\usepackage[latin1]{inputenc}
\usepackage[T1]{fontenc}
\usepackage{amsmath}
\usepackage{amsfonts}
\usepackage{graphicx}
\usepackage{subfigure}
\usepackage{a4wide}
\usepackage{amssymb}
\usepackage{fancyhdr}
\usepackage{mathrsfs}
\usepackage[toc,page]{appendix}

\begin{document}

\title{\bf Inhomogeneous viscous fluids in FRW universe and finite-future time singularities}
\author{ 
Shynaray Myrzakul\footnote{Email: shynaray1981@gmail.com},\,\,\,
Ratbay Myrzakulov\footnote{Email: rmyrzakulov@gmail.com},\,\,\,
Lorenzo Sebastiani\footnote{E-mail address: l.sebastiani@science.unitn.it
}\\
\\
\begin{small}
Eurasian International Center for Theoretical Physics and  Department of General
\end{small}\\
\begin{small} 
Theoretical Physics, Eurasian National University, Astana 010008, Kazakhstan
\end{small}\\
}

\date{}

\maketitle


\begin{abstract}

We consider inhomogeneous viscous fluids in flat Friedmann-Robertson-Walker universe.
We analyze different kinds of such fluids and investigate the possibility to reproduce the current cosmic acceleration providing a different future evolution with respect to the Cosmological Constant case. In particular, we study the presence of finite-future time singularities. We also discuss a general class of ``integrable'' viscous fluid models whose bulk viscosities obey to a common differential equation.
\end{abstract}



\tableofcontents
\section{Introduction}

Since the discovery of the cosmic accelerated expansion of the universe~\cite{SN11}--\cite{WMAP}, 
several descriptions of the so called dark energy issue have been presented.
The simplest one is the introduction of small and positive
Cosmological Constant in the framework of General Relativity ($\Lambda$CDM Model). In this case, the acceleration is the effect of the negative pressure of the dark energy fluid with Equation of State parameter 
$\omega=-1$.
Today, the nature of the dark energy has become the ``Mystery of the Millennium''~\cite{Pad} and is probably the most ambitious and tantalizing field of research because of its implications in fundamental physics. If from one side that the dark fluid has an Equation of State parameter very close to minus one represents an important point in favour of the Cosmological Constant-like representation, the estimated extremely small value of Cosmological Constant leads to several well-know problems and the idea according to which General Relativity plus Cosmological Constant is not the ultimate theory of gravity, but an extremely good approximation valid in the present day range of detection, is well accepted.
Moreover, the last cosmological data constrain $\omega$ of the dark energy as $\omega=-0.972^{+0.061}_{-0.060}$, so that
different
forms of dark fluid (phantom, quintessence, inhomogeneous fluids, \textit{etc}.), satisfying a suitable Equation of State are allowed. The study of non-perfect fluids in the Friedmann-Robertson-Walker universe can be justified by several argumentations. First of all, 
despite to the fact that many macroscopic physical systems, like the large scale structure of matter and radiation in the universe, can be approximated like perfect fluids (with Equation of State $p=\omega\rho$, $\omega$ constant), we cannot exclude a different composition for the dark energy, whose origin remains unknown. Then, in the last years, the interest in modified theories of gravity has grown up. These theories imply a modification to Einstein's gravity and  some combination of curvature invariants (Riemann tensor, Weyl tensor, Ricci tensor and so on) replaces or is added into the classical Hilbert-Einstein action of General Relativity. In this way, the current acceleration is produced by some (sub)-dominant terms of gravitational action which become essential at small curvatures
(see Refs.~\cite{Review-Nojiri-Odintsov} --\cite{plus} for review on $F(R)$-gravity and 
Ref.\cite{others} for recent works). 
It is worth considering that the modified theories of gravity have a corresponding description in the fluid-like form, and the study of inhomogeneous viscous fluids is one of the easiest way to understand some general features of such a kind of alternative theories.

In this paper we would like to analyze different kinds of inhomogeneous viscous fluids in flat Friedmann-Robertson-Walker space-time. We investigate the possibility to reproduce the current cosmic acceleration providing different future evolutions with respect to the Cosmological Constant case. In particular, we are interested in the presence of finite-future time singularities, namely, a finite-time where the scale factor, the energy density, the pressure, the Hubble paramter or more simply some derivatives of Hubble paramter diverge. It is well-know that such a scenario is realized by phantom perfect fluids (Big Rip), but by starting from more general forms of fluid, other possibilities are permitted.

The paper is organized as follows. In Section {\bf 2}, we will revisit the equations for Friedmann-Robertson-Walker universe and the feature of finite-future time singualarity solutions. Moreover, we will introduce the formalism of inhomogeneous viscous fluids.
In Sections {\bf 3}--{\bf 4}, we will analyze viscous fluids with constant and not constant EoS parameter, separately. In the specific, we will investigate the possibility to reproduce the current cosmic accelereation according with observations (in particular, the effective EoS paramter of the fluid must be close to minus one) with a singularity scenario in the future evolution of the universe. In Section {\bf 5}, we will discuss a class of ``integrable'' viscous fluid models whose viscosities obey to a common integrable equation. Conclusions are given in Section {\bf 6}.

We use units of $k_{\mathrm{B}} = c = \hbar = 1$ and denote the
gravitational constant, $G_N$, by $\kappa^2\equiv 8 \pi G_{N}$, such that
$G_{N}^{-1/2} =M_{\mathrm{Pl}}$, $M_{\mathrm{Pl}} =1.2 \times 10^{19}$ GeV being the Planck mass.


\section{Fluids in flat FRW space-time and finite-future time singularities}

The Friedmann equations for the flat Friedmann-Robertson-Walker (FRW) metric 
\begin{equation}
ds^{2}=-dt^{2}+a^{2}(t)d\mathbf{x}^{2}\,,
\end{equation}
$a(t)$ being the scale factor of the universe, read
\begin{equation}
\frac{3H^2}{\kappa^2}=\rho\,,\quad\frac{2\dot H}{\kappa^2}=-(\rho+p)\,,\label{EOMs}
\end{equation}
where $H=\dot a(t)/a(t)$ is the Hubble paramter and the dot denotes the derivative with respect to the cosmological time. Moreover, $p$ and $\rho$ are the pressure and the energy density of the fluid contents of the universe. 
From this equations one derives the conservation law,
\begin{equation}
\dot\rho+3H(\rho+p)=0\,.\label{CL}
\end{equation} 
For perfect fluids the Equation of State (EoS)
is expressed as
\begin{equation}
p=\omega\rho\,,
\end{equation}
where $\omega$ is the constant EoS paramter (for example, for matter $\omega=0$ and for radiation $\omega=1/3$). In this case, the solutions of Eqs.~(\ref{EOMs}) read
\begin{equation}
a(t)=a_0(t+t_0)^{\frac{2}{3(1+\omega)}}\,,\quad H(t)=\frac{2}{3(1+\omega)(t+t_0)}\,,\quad\rho(t)=\rho_0a(t)^{-3(1+\omega)}\,,\quad\omega\neq-1\,,
\end{equation}
$a_0\,,\rho_0\,,t_0$ being integration constants eventually related to each other.
When $\omega=-1$ (it corresponds to the Cosmological Constant case), one has
\begin{equation}
a(t)=a_0 \text{e}^{\sqrt{\kappa^2\rho_0/3} t}\,,\quad H=\sqrt{\frac{\kappa^2\rho_0}{3}}\,,
\quad\rho=\rho_0\,,
\quad\omega=-1\,.
\end{equation}
It is easy to verify that $\ddot a>0$ when $-1/3>\omega$, which corresponds to the violation of the strong energy condition (SEC). We get three types of perfect fluids giving acceleration: quintessence ($-1/3>\omega>-1$), whose energy density decreases with the expansion; Cosmological Constant ($\omega=-1$), whose corresponding energy density is constant with the expansion;
phantom ($-1>\omega$), whose energy density increases with the expansion.

In expanding universe ($H>0$), if $\omega>-1$, $t_0>0$ and usually one puts $t_0=0$. On the other hand, if $-1>\omega$ (phantom fluids), in order to have an expansion we must require $0>t_0$ and $-t_0>t$. In this case, we can rewrite the Hubble paramter as
\begin{equation}
H(t)=-\frac{2}{3(1+\omega)(t_0-t)}\,,\quad t<t_0\,,
\end{equation}
where we have shifted $t_0\rightarrow-t_0>0$. As a consequence, when $t$ is close to $t_0$,
the Hubble paramter, the energy density and the scale factor diverge and the universe becomes singular at the finite time $t_0$, namely the Big Rip scenario occurs~\cite{Caldwell}.

\subsection{Finite-future time singularities}

If we consider non perfect fluids
in FRW universe,
other kinds of finite-future time singularities different from the Big Rip
may appear as solutions of Friedmann equations.

We expect a finite-future time singularity in expanding universe
when the Hubble parameter assumes the form
\begin{equation}
H(t)=\frac{h_0}{(t_{0}-t)^{\beta}}+H_{0}\,,\quad t<t_0\,,\beta\neq 0\,,
\label{Hsingular}
\end{equation}
where $h_0$, $t_{0}$ and $H_{0}$ are positive constants,
$\beta$ is a generic parameter which describes the type of singularity and $t_0$ is the finite-time for which singularity is realized.
We exclude the case $\beta=0$, which corresponds to de Sitter space.

The finite-future time singularities can be classified
as~\cite{classificationSingularities}:
\begin{itemize}
\item Type I (Big Rip \cite{Rip1}-\cite{Rip19}): for $t\rightarrow t_{0}$, $a(t)\rightarrow\infty$,
$\rho\rightarrow\infty$ and
$|p|\rightarrow\infty$.
It corresponds to the cases $\beta=1$ and $\beta>1$.
\item Type II (sudden \cite{s1,sudden, s2}):
for $t\rightarrow t_{0}$, $a(t)\rightarrow a_{0}$,
$\rho\rightarrow\rho_{0}$ and $|p|
\rightarrow\infty$.
It corresponds to the case $-1<\beta<0$.
\item Type III (introduced in Ref.~\cite{classificationSingularities}): for $t\rightarrow t_{0}$, $a(t)\rightarrow a_{0}$,
$\rho\rightarrow\infty$ and
$|p|\rightarrow\infty$.
It corresponds to the case $0<\beta<1$.
\item Type IV (introduced in Ref.~\cite{classificationSingularities}): for $t\rightarrow t_{0}$, $a(t)\rightarrow a_{0}$,
$\rho\rightarrow \rho_0$, $|p|
\rightarrow p_0$
and higher derivatives of $H$ diverge.
It corresponds to the case
$\beta<-1$, but $\beta$ is not any integer number.
\end{itemize}
Here, $a_{0} (\neq 0)$ and $\rho_{0}$, $p_{0}$ are constants.
We call singularities for $\beta=1$ and
those for $\beta>1$ as the `Big Rip' singularities and the `Type I'
singularities, respectively.

The scale factor corresponding to the four types of singularity behaves as
\begin{eqnarray}
\left\{\begin{array}{lll}
a(t)&=& \frac{a_0}{(t_0-t)^h}\,,\quad\quad\quad\quad\quad\quad\quad\quad\quad\beta=1\quad\text{(Big Rip)}\,;\\ \\
a(t)&=& a_0\exp\left[\frac{h_0(t_0-t)^{1-\beta}}{\beta-1}\right]\,,\quad\beta(\neq 1)>0\quad\text{(Type I, III singularities)}\,;\\ \\
a(t)&=& a_0\exp\left[\frac{h_0(t_0-t)^{1-\beta}}{\beta-1}+H_0\,t\right]\,,\quad\beta<0\quad\text{(Type II, IV singularities)}\label{asingularforms}\,.
\end{array}\right.
\end{eqnarray} 
Here, $H_0\neq 0$ has been considered in the significant cases of Type II and IV singularities only.

Singularity solutions are interesting to investigate: as in the case of the Big Rip for the phantom perfect fluid, they may represent a possible future evolution scenario for our universe admitting the acceleration at the present epoch, since when $(t/t_0)\ll 1$ the Hubble parameter in (\ref{Hsingular}) reads 
\begin{equation}
H(t\ll t_0)\simeq\frac{h_0}{t_0^\beta}+H_0
+\mathcal{O}\left(t/t_0\right)\,,
\end{equation}
such that at the present time $H\simeq H^0$ we must require
\begin{equation}
H^{0}\simeq\frac{h_0}{t_0^\beta}+H_0\,,
\end{equation}
where $H^{0}\simeq 10^{-66}eV^{-2}$ is the (nearly constant) value of the Hubble parameter in the universe today. For phantom perfect fluids, $\beta=1$ and $h_0=-2/(3(1+\omega))$.

\subsection{Inhomogeneous viscous fluids in flat FRW space-time}

The equation of state of inhomogeneous viscous fluids in a flat Friedmann-Robertson-Walker space-time 
is given by~\cite{fluidsOd2, Od3, Od4}
\begin{equation}
p=\omega(\rho)\rho+B(\rho,a(t),H, \dot{H}...)\,,\label{start}
\end{equation}
where the EoS parameter, $\omega(\rho)$, may depend on the energy density, and the bulk viscosity  $B(\rho,a(t),H, \dot{H}...)$ is a general function of the fluid energy density, the scale factor, the Hubble parameter and its derivatives. 
In this paper we will consider a simple formulation of such equation, namely
\begin{equation}
p=\omega(\rho)\rho-3 H\zeta(H)\,,\label{eq.state}
\end{equation}
where $\zeta(H)$ is the bulk viscosity and it depends on the Hubble parameter $H$ only. On thermodynamical grounds, in order to have the positive sign of the entropy change in an irreversible process, $\zeta(H)$ has to be a positive quantity, so that we assume $\zeta(H)>0$~\cite{Alessia, Alessia(2)}. 
Moreover, for the stress-energy tensor of fluid $T_{\mu\nu}$ one has
\begin{equation}
T_{\mu\nu}=\rho u_{\mu}u_{\nu}+\left[\omega(\rho)\rho-3H\zeta(H)\right](g_{\mu\nu}+u_{\mu}u_{\nu})\,, 
\end{equation}
where $u_{\mu}=(1,0,0,0)$ is the four velocity vector. 
The fluid energy conservation law reads
\begin{equation}
\dot{\rho}+3H\rho(1+\omega(\rho))=9H^{2}\zeta(H) \label{conservationlawfluid}\,.
\end{equation}
In what follows, we will analyze the behaviour of such a kind of fluids in FRW universe. We are interested in fluids which provide a viable cosmology today but a different future evolution with respect to the Cosmological Constant case. In the specific, we will investigate the possibility to recover singularity solutions as in Eq.~(\ref{Hsingular}), which lead to the de Sitter space for small value of $t$.

We remind that the effective EoS paramter for fluid (\ref{eq.state}) reads
\begin{equation}
\omega_{\text{eff}}:=\frac{p}{\rho}=\omega(\rho)-\frac{3H\zeta(H)}{\rho}\,.\label{eff}
\end{equation}
Cosmological data imply $\omega_{\text{eff}}\simeq -1$. In the case of perfect fluids, it follows that $\omega$ must be very close to the value of minus one, 
but for different kinds of non perfect fluid other possibilities are allowed.

We will consider the cases of $\omega$ constant and $\omega(\rho)$ not a constant separatley. 

\section{Viscous fluids with $\omega$ constant}

We start by considering $\omega(\rho)=\omega$ constant and different forms of bulk viscosity (see also Refs.~\cite{B1,B2,B3} for applications of such a kind of fluids to early time acceleration).

\subsection{Constant viscosity\label{3.1}}

Let us remind the case of constant bulk viscosity  $\zeta(H)=\zeta_0$, $\zeta_0>0$. Eq.~(\ref{conservationlawfluid}) yields
\begin{equation}
\rho=\rho_{0}a^{-3(1+\omega)}+9\zeta_{0} a^{-3(1+\omega)}\int^{t} a(t')^{1+3\omega}\dot{a}(t')^{2}dt'\,,
\end{equation}
where $\rho_0$ is an integration constant.
By combining the first Friedmann equation with the conservation law, we get~\cite{Alessia}
\begin{equation}
\rho=\frac{9\rho_{0}}
{\left[3+\left[\frac{3}{\zeta_0}(1+\omega)\sqrt{\frac{\rho_0}{3\kappa^2}}(1-\mathrm{e}^{-(3\kappa^2\zeta_{0}/2)t})\right]\right]^2}\,,
\quad
H(t)=\frac{\sqrt{3\kappa^2\rho_{0}}}{3+\left[\frac{3}{\zeta_0}(1+\omega)\sqrt{\frac{\rho_0}{3\kappa^2}}(1-\mathrm{e}^{-(3\kappa^2\zeta_{0}/2)t})\right]}\,.\label{exp}
\end{equation}
If we recast in the above expression
\begin{equation}
t_0=-\frac{2}{3\kappa^2\zeta_0}\ln\left[1+\sqrt{\frac{3\kappa^2}{\rho_0}}\frac{\zeta_0}{(1+\omega)}\right]\,,
\end{equation}
we can rewrite the Hubble parameter as
\begin{equation}
H=-\frac{\zeta_0}{(1+\omega)}\frac{\text{e}^{\frac{3\kappa^2\zeta_0\,t_0}{2}}\kappa^2}
{\left(\text{e}^{\frac{3\kappa^2\zeta_0(t_0-t)}{2}}-1\right)}\,.
\end{equation}
We immediatly see that, if $\omega<-1$, $H$ is positive when $0<t<t_0$ and diverges at $t\rightarrow t_0$. On the other hand, if $\omega>-1$, we must require $t>t_0$ and the divergence is removed. In such a case, by putting $t_0=0$, we derive
\begin{equation}
\frac{\ddot a}{a}=H^2+\dot H=
\frac{\zeta_0^2(\kappa^2)^2}{(1+\omega)}\frac{1}
{\left(1-\text{e}^{\frac{-3\kappa^2\zeta_0\,t}{2}}\right)^2}
\left[
\frac{1}{(1+\omega)}
-\frac{3\text{e}^{\frac{-3\kappa^2\zeta_0\,t}{2}}}{2}
\right]
\,,
\end{equation}
such that for $\omega<-1/3$ the fluid produces an accelerated expansion when $t>0$.

If $\omega<-1$, it is possible to see that the singularity which emerges at $t=t_0$ corresponds to the Big Rip one. As a matter of fact, for
$t\rightarrow t_{0}$,
we may expand $t$ around $t_{0}$ obtaining
\begin{equation}
H(t)\simeq-\frac{2}{3(1+\omega)}\frac{1}{(t_{0}-t)}
+\mathcal{O}(t_{0}-t)\,,\label{oraetlaborabis}
\end{equation}
which corresponds to the case $\beta=1$ in (\ref{Hsingular}). Moreover, in the limit $(t/t_0)\ll 1$\,,
\begin{equation}
H(t)\simeq
-\frac{\zeta_0\kappa^2}{(1+\omega)}\frac{1}{\left(1-\text{e}^\frac{-3\kappa^2\zeta_0 t_0}{2}\right)}+\mathcal{O}\left(t/t_0\right)\,,\label{dsu}
\end{equation}
and we get a (quasi) de Sitter solution. 
The effective EoS paramter (\ref{eff}) of the fluid is given by
\begin{equation}
\omega_{\text{eff}}=\omega-\frac{3H\zeta_0}{\rho}\,,
\end{equation}
which reads for (\ref{dsu})
\begin{equation}
\omega_{\text{eff}}\simeq 2\omega+1\,,
\end{equation}
where we have taken in account that $0\ll t_0$. This model (with $\omega<-1$) may be compatible with cosmological data if $\omega$ is close to the value of $-1$ and $\zeta_0\simeq -H^0 (1+\omega)/\kappa^2$, being $H^0$ the Hubble paramter of the universe today. It follows that the bulk viscosity must be small.

\subsection{Viscosity proportional to $H$\label{3.2}}

This is the case $\zeta(H)=3H\tau$. Since $\zeta$ must be positive, the constant $\tau$ is also assumed to be positive. 
Eq.~(\ref{conservationlawfluid}) reads
\begin{equation}
\rho=\rho_{0}a^{-3(1+\omega_{\mathrm{F}})}+27\tau a^{-3(1+\omega)}\int^{t} dt'a(t')^{3\omega}\dot{a}(t')^{3}\,.
\end{equation}
The solution of Friedmann equations and energy conservation law lead to 
\begin{equation}
\rho=\frac{4}{3\gamma^2\kappa^2}\frac{1}{(t_0-t)^2}\,,
\quad
H(t)=\left(\frac{2}{3\gamma}\right)\frac{1}{(t_0-t)}\,,\label{HHH}
\end{equation}
where
\begin{equation}
\gamma=-(1+\omega-3\tau\kappa^2)\,.
\end{equation}
Since $\gamma(t_0-t)$ has to be positive in expanding universe, when $\gamma<0$, we can set $t_0=0$ and get a solution for accelerating universe, namely
\begin{equation}
H(t)=-\left(\frac{2}{3\gamma}\right)\frac{1}{t}>0\,,\quad\frac{\ddot a}{a}=
\left(\frac{2}{3\gamma}\right)\frac{1}{t^2}\left[\left(\frac{2}{3\gamma}\right)+1\right]>0\,,
\end{equation}
under the condition
\begin{equation}
-\frac{2}{3}<\gamma<0\quad\Rightarrow\quad 0<(1+\omega-3\tau\kappa^2)<\frac{2}{3}\,.
\end{equation}
On the other hand, our fluid realizes the Big Rip scenario with
$h_0=2/(3\gamma)$ when $\gamma>0$, namely~\cite{Alessia},
\begin{equation}
(1+\omega)-3\kappa^2\tau<0\,.\label{zizizip}
\end{equation}
In this case, $t_0>0$ in (\ref{HHH}) is the time for which singularity emerges.
Phantom fluids or quintessence fluids with sufficiently large bulk viscosity could produce this kind of finite-future time singularity. The effective EoS paramter reads 
\begin{equation}
\omega_{\text{eff}}=\omega-\frac{9H^2\tau}{\rho}\,.
\end{equation}
When $(t/t_0)\ll 1$, the Hubble paramter is approximated as
\begin{equation}
H(t\ll t_0)\simeq\frac{2}{3\gamma t_0}
+\mathcal{O}\left(t/t_0\right)\,,
\end{equation}
such that
\begin{equation}
\omega_{\text{eff}}\simeq \omega-3\tau\kappa^2\,,
\end{equation}
and $\omega$ has to be $\omega\simeq-1+3\tau\kappa^2$ to recover $\omega_{\text{eff}}\simeq-1$.
This kind of fluid could have the $\omega$ paramter arbitrarily large (and eventually positive) compatibly with the (positive) bulk viscosity $\tau$, which guarantees the violation of the strong energy condition ($\omega_{\text{eff}}<-1/3$). It follows that in order to reproduce the universe today ($1/t_0\simeq 3H^0\gamma/2$), the singularity time $t_0$ has to be large since $0<\gamma\ll 1$.

\subsection{Viscosity proportional to $H^n$\label{3.3}}

This is the case $\zeta(H)=(3H)^n\tau$, where $n$ is a free parameter and $\tau$ is positive. 
The conservation law leads to
\begin{equation}
\rho=\rho_{0}a^{-3(1+\omega_{\mathrm{F}})}
+3^{n+2}\tau a^{-3(1+\omega)}\int^{t} dt'a(t')^{(1-n)+3\omega}\dot{a}(t')^{2+n}\,.
\end{equation}
The simplest case is given by $\omega=-1$, for which the solutions of Friedmann equations and energy conservation law are given by 
\begin{equation}
\rho=\frac{1}{3\kappa^2}\left(\frac{2}{3n\tau\kappa^2}\right)^{\frac{2}{n}}
\frac{1}{(t_0-t)^{\frac{2}{n}}}\,,\quad
H=\frac{1}{3}\left(\frac{2}{3n\kappa^2\tau}\right)^{\frac{1}{n}}
\frac{1}{(t_0-t)^{\frac{1}{n}}}\,.
\end{equation}
Since $\tau$ is positive defined, when $n<0$ we obtain (by setting $t_0=0$)
\begin{equation}
\rho=\frac{1}{3\kappa^2}\left(-\frac{2}{3n\tau\kappa^2}\right)^{\frac{2}{n}}
(t)^{-\frac{2}{n}}\,,\quad
H=\frac{1}{3}\left(-\frac{2}{3n\kappa^2\tau}\right)^{\frac{1}{n}}
(t)^{-\frac{1}{n}}\,,\quad n<0\,,
\end{equation}
such that we have an acceleration when
\begin{equation}
\frac{\ddot a(t)}{a(t)}=\frac{1}{3n}\left(-\frac{2}{3n\kappa^2\tau}\right)^{\frac{1}{n}}t^{-\frac{(n+1)}{n}}\left[\frac{1}{3}\left(-\frac{2}{3n\kappa^2\tau}\right)^{\frac{1}{n}}n t^{\frac{n-1}{n}}-1\right]>0\,,\quad n<0\,.
\end{equation}
This expression may effectively lead to an acceleration for $t>0$.
Otherwise, when $n>0$, a finite-time singularity appears in the future universe scenario, namely, a Type I singularity when $0<n<1$, and a Type III singularity when $1<n$. Note that $n=1$ corresponds to the special case $\omega=-1$ of the previous Subsection. The effective EoS paramter is derived as
\begin{equation}
\omega_{\text{eff}}=-1-\frac{(3H)^{n+1}\tau}{\rho}\,.
\end{equation}
Since when $(t/t_0)\ll 1$, the Hubble paramter reads
\begin{equation}
H(t\ll t_0)\simeq\frac{1}{3}\left(\frac{2}{3n\kappa^2\tau\,t_0}\right)^{\frac{1}{n}}
+\mathcal{O}\left(t/t_0\right)\,,
\end{equation}
the effective EoS paramter of the viscous fluid related to this solution results to be 
\begin{equation}
\omega_{\text{eff}}\simeq -1-3\left(\frac{2}{3n\,t_0}\right)^\frac{n-1}{n}(\kappa^2\tau)^{\frac{1}{n}}\,.
\end{equation}
This expression is close to minus one for small values of bulk viscosity $\tau$ and as a consequence, in order to reproduce the universe today ($1/t_0\simeq (3H^0)^n (3n\kappa^2\tau/2)$), the singularity time $t_0$ must be large.\\
\\
An other interesting case for which we obtain an explicit solution of the EOMs when $\zeta(H)=(3H)^n\tau$, is given by the special choice $n=-1$, $\omega$ generic EoS paramter. We get 
\begin{equation}
\rho=\frac{\tau}{(1+\omega)}\tanh^2\left[\frac{\sqrt{3\kappa^2\tau(1+\omega)}}{2}(t-T)\right]\,,
H=\sqrt{\frac{\tau\kappa^2}{3(1+\omega)}\tanh^2\left[\frac{\sqrt{3\kappa^2\tau(1+\omega)}}{2}(t-T)\right]}\,.\label{solsol}
\end{equation}
Here, $T$ is an integration constant. Since $\tau>0$, it follows that when $\omega>-1$, we can put $T=0$ and obtain a solution for accelerated expanding universe with
\begin{equation}
\frac{\ddot a(t)}{a(t)}=\frac{\tau\kappa^2}{3(1+\omega)}
\tanh^2\left[\frac{\sqrt{3\kappa^2\tau(1+\omega)}}{2} t \right]+\frac{\tau\kappa^2}{2}
\cosh^{-2}
\left[\frac{\sqrt{3\kappa^2\tau(1+\omega)}}{2} t\right]>0\,.
\end{equation}
On the other hand, if $\omega<-1$, we may rewrite (\ref{solsol}) as
\begin{equation}
\rho=-\frac{\tau}{(1+\omega)}\tan^2\left[\frac{\sqrt{3\kappa^2\tau |1+\omega|}}{2}(t-T)\right]\,,
H=\sqrt{-\frac{\tau\kappa^2}{3(1+\omega)}\tan^2\left[\frac{\sqrt{3\kappa^2\tau|1+\omega|}}{2}(t-T)\right]}\,.
\end{equation}
This is a solution for expanding universe with $T<t$ which possesses a future-finite time singularity located at
\begin{equation}
t_0=T+\frac{\pi}{\sqrt{3\kappa^2\tau|1+\omega|}}\,.
\end{equation}
Here, $T$ is considered to be such that $0<T<\pi/\sqrt{3\kappa^2\tau|1+\omega|}$, and the argument of the tangent changes between $T$ and $\pi/2$ when $t$ changes between $0$ and $t_0$.
When $t$ is close to $t_0$, the Hubble paramter reads
\begin{equation}
H\simeq-\frac{2}{3(1+\omega)}\frac{1}{(t_0-t)}+\mathcal{O}(t_{0}-t)\,,
\end{equation}
and we identify the singularity as the Big Rip one. On the other hand, when $(t/t_0)\ll 1$, the Hubble paramter is given by
\begin{equation}
H(t\ll t_0)\simeq
\sqrt{-\frac{\tau\kappa^2}{3(1+\omega)}\tan^2\left[\frac{\sqrt{3\kappa^2\tau|1+\omega|}}{2}(t_0)-\frac{\pi}{2}\right]}+\mathcal{O}(t/t_0)\,.\label{star}
\end{equation}
Furthermore, the effective EoS paramter is derived as
\begin{equation}
\omega_{\text{eff}}=\omega-\frac{\tau}{\rho}\,,
\end{equation}
which corresponds to
\begin{equation}
\omega_{\text{eff}}\simeq
\omega+(1+\omega)\Big\vert\tan^{-1}\left[\frac{\sqrt{3\kappa^2\tau|1+\omega|}}{2}(t_0)-\frac{\pi}{2}\right]\Big\vert\,,
\end{equation}
in the case of (\ref{star}). As a consequence, in order to reproduce the nearly constant value of the dark energy in our universe, we must require $\omega$ close to minus one, such that $\omega_\text{eff}\simeq-1$. 
However, in such a case, $\sqrt{\tau}\,t_0$ must be large enough
to guarantee 
$H\simeq H^0$, $H^0$ being the Hubble paramter of the universe today.

\section{Inhomogeneous viscous fluids with $\omega(\rho)$ not a constant}

In this Section, we will take $\omega(\rho)$ explicitly depending on the energy density of the fluid. 
In particular, we will consider the following simple form of $\omega(\rho)$~\cite{NO1,NO2},
\begin{equation}
\omega(\rho)=A_{0}\rho^{\alpha-1}-1\label{omega}\,,
\end{equation}
where $A_{0}$($\neq 0$) and $\alpha$($\neq 1$) are constants. 
By considering the bulk viscosity as $\zeta(H)=(3 H)^{n}\tau$, $\tau\,, n>0$ again, we get
the energy conservation law 
\begin{equation}
\dot{\rho}+3 H A_{0}\rho^{\alpha}=9H^{2}(3H)^{n}\tau\label{super}\,.
\end{equation}
We start with the simplest case $\tau=0$ (non viscous case). The solutions of the Friedmann equations result to be
\begin{equation}
\rho=\left[-
\frac{2}{\sqrt{3\kappa^2}(2\alpha-1)A_0}
\right]^\frac{2}{2\alpha-1}\frac{1}{(t_0-t)^{\frac{2}{2\alpha-1}}}
\,,\quad 
H=\sqrt{\frac{\kappa^2}{3}}
\left[-
\frac{2}{\sqrt{3\kappa^2}(2\alpha-1)A_0}
\right]^\frac{1}{2\alpha-1}\frac{1}{(t_0-t)^{\frac{1}{2\alpha-1}}}\,.
\end{equation}
It follows that, if $A_0(2\alpha-1)>0$, we may put $t_0=0$ obtaining a solution for expanding universe with
\begin{equation}
\frac{\ddot a}{a}=
\frac{(\kappa^2)^\frac{\alpha-1}{2\alpha-1}}{3^\frac{\alpha}{2\alpha-1}}
\left(\frac{2}{A_0}\right)^{\frac{1}{2\alpha-1}}
\frac{1}{(2\alpha-1)^\frac{2\alpha}{2\alpha-1}}
\frac{1}{t^\frac{2\alpha}{2\alpha-1}}
\left[
\frac{(\kappa^2)^\frac{\alpha-1}{2\alpha-1}}{3^\frac{\alpha}{2\alpha-1}}
\left(\frac{2}{A_0}\right)^{\frac{1}{2\alpha-1}}
\frac{1}{(2\alpha-1)^\frac{2-2\alpha}{2\alpha-1}}
\frac{1}{t^\frac{2-2\alpha}{2\alpha-1}}-1
\right]\,,
\end{equation}
such that acceleration is allowed for $t>0$. On the other hand, if 
$A_0(2\alpha-1)<0$, a finite-future time singularity appears at $t=t_0$.
In such a case, several different types of singularities may be realized.
Namely, if $1/2<\alpha<1$, a Type I singularity appears; if $1<\alpha$, a Type III singularity appears; if $0<\alpha<1/2$, a Type IV singularity appears; finally, if $\alpha<0$, a Type II singularity appears (see Refs.~\cite{NO1, NO2} where these cases have firstly been analyzed for that fluid).
The effective EoS paramter of the fluid results to be
\begin{equation}
\omega_{\text{eff}}=A_0\rho^{\alpha-1}-1\,.
\end{equation}
When $(t/t_0)\ll 1$, we may easily evaluate from 
\begin{equation}
H(t\ll t_0)\simeq
\sqrt{\frac{\kappa^2}{3}}
\left[-
\frac{2}{\sqrt{3\kappa^2}(2\alpha-1)A_0}
\right]^\frac{1}{2\alpha-1}\frac{1}{t_0^{\frac{1}{2\alpha-1}}}+\mathcal{O}(t/t_0)\,,
\end{equation}
the value of the effective EoS paramter as
\begin{equation}
\omega_\text{eff}\simeq A_0^{\frac{1}{2\alpha-1}}\left[-
\frac{2}{\sqrt{3\kappa^2}(2\alpha-1)}
\right]^\frac{2\alpha-2}{2\alpha-1}\frac{1}{(t_0)^{\frac{2\alpha-2}{2\alpha-1}}}-1\,.\label{tic}
\end{equation}
In order to obtain $\omega_\text{eff}\simeq -1$
in this kind of fluid with $A_0(2\alpha-1)<0$, we must require $A_0\gg 1$ if $0<A_0$ and $(2\alpha-1)<0$, and $|A_0|\ll 1$ if $A_0<0$ and $0<(2\alpha-1)$. In the both cases, we are in the quintessence region and
in order to reproduce the universe today with $H\simeq H^0$ the singularity time $t_0$ must be large.

Other interesting cases of non viscous inhomogeneous fluids reproducing viable cosmology have been presented in Ref.~\cite{NO2}. Here, the EoS paramter is in the form of $\omega{\rho}=f(\rho)-1$, where $f(\rho)$ is a function of the fluid energy density. It is interesting to note that the authors present some examples of fluids in the phantom region without finite-future time singularity. The fact is that the energy density grows with time but not rapidly enough for the occurrence of the singularity and we have the so-called `Little Rip'~\cite{LR} and a dissolution of bound structures at some point in the future may occur. 
\\
\\
As a second case, let us take $\tau\neq 0$ (viscous case) and consider $n=2\alpha-1$, for which the Friedmann equations lead to
\begin{equation}
\rho=
\gamma
^\frac{2}{1-2\alpha}
\frac{1}{\left(t_0-t\right)^\frac{2}{2\alpha-1}}\,,\quad
H=
\sqrt{\frac{\kappa^2}{3}}
\gamma
^\frac{1}{1-2\alpha}
\frac{1}{\left(t_0-t\right)^\frac{1}{2\alpha-1}}\,,
\end{equation}
where
\begin{equation}
\gamma=\left[\left[3^\frac{2\alpha+1}{2}(\kappa^2)^\frac{2\alpha+1}{2}\tau-A_0\sqrt{3\kappa^2}
\right]
\left[
\frac{2\alpha-1}{2}
\right]\right]\,.\label{gg}
\end{equation}
If $\gamma<0$, we may put $t_0=0$ obtaining an expansion with
\begin{equation}
\frac{\ddot a}{a}=
\sqrt{\frac{\kappa^2}{3}}
\frac{(-\gamma)
^\frac{1}{1-2\alpha}}
{(2\alpha-1)}
\frac{1}{t^\frac{2\alpha}{2\alpha-1}}
\left[\sqrt{\frac{\kappa^2}{3}}\frac{(-\gamma)^{\frac{1}{1-2\alpha}}(2\alpha-1)}{t^{\frac{2-2\alpha}{2\alpha-1}}}
-1\right]\,,\quad\gamma<0\,.
\end{equation}
Acceleration is allowed for $t>0$ if $(2\alpha-1)>0$. Otherwise, when $\gamma>0$, a finite-future time singularity appears at $t=t_0$, $t_0>0$. As in the previous non viscous case, all the four types of singularities can be realized by the fluid, whose effective EoS paramter reads now
\begin{equation}
\omega_{\text{eff}}=A_0\rho^{\alpha-1}-1-\frac{(3H)^{2\alpha}\tau}{\rho}\,.
\end{equation}
Since for $(t/t_0)\ll 1$, 
\begin{equation}
H(t\ll t_0)\simeq
\sqrt{\frac{\kappa^2}{3}}
\gamma^\frac{1}{1-2\alpha}\frac{1}{t_0^{\frac{1}{2\alpha-1}}}+\mathcal{O}(t/t_0)\,,
\end{equation}
we derive
\begin{equation}
\omega_\text{eff}\simeq 
-1+\frac{A_0\gamma^{\frac{2\alpha-2}{1-2\alpha}}}{t_0^{\frac{2\alpha-2}{2\alpha-1}}}
-\frac{(3\kappa^2)^\alpha\tau}{(\gamma\,t_0)^{\frac{2\alpha-2}{2\alpha-1}}}\,.
\end{equation}
By putting $\tau=0$ and taking into account (\ref{gg}), we recover (\ref{tic}).
In order to get $\omega_\text{eff}\simeq -1$ and to reproduce the current value of the Hubble paramter when $\tau\neq 0$, it is enough to require $|A_0\,,\tau|\ll 1$ and $t_0\gg 0$.\\
\\
We would like to conclude this Section with some comments about a special non-viscous fluid with non constant EoS paramter, namely the Chaplygin gas~\cite{Ciappinski}, which has been studied as a possible candidate for the dark energy in many works. The Equation of State of Chaplygin gas reads
\begin{equation}
p=-\frac{A_{0}}{\rho}\,,
\end{equation}
where $A_{0}$ is a positive constant. 
When $\omega_{\text{eff}}=-A_0/\rho^2<-1/3$, this gas produces an accelerated expansion.
The energy conservation law leads to\\
\phantom{line}
\begin{equation}
\rho_{\mathrm{F}}=\sqrt{A_{0}+\frac{1}{a(t)^{6}}}\label{rhoChap}\,.
\end{equation}
\phantom{line}\\
Since when $a(t)$ diverges, the energy density (and therefore the pressure) tends to be constant, Chaplygin gas cannot realize a singularity scenario in the future evolution of the universe.

\section{Integrable fluid models in FRW universe}

Let us return to the FRW equation system (\ref{EOMs}) with inhomogeneous  viscous fluid (\ref{eq.state}). In principle, to solve these equations, we need the explicit EoS of the fluid, as we did in the previous sections. Here, we would like to consider an other approach to viscous fluids, namely the so called integrable viscous fluid models (see Refs. \cite{Integrable}--\cite{Abis} and references therein), where the bulk viscosity obeys to some specific  equation.
The idea is the following. 
If one parameter of the fluid, say the bulk viscosity $\zeta(H)$ for our case, satisfies some integrable differential equation, we obtain an ``integrable FRW model '' which belongs to a general class of integrable models. In this way, we can study the feauture of different models whose bulk viscosities satisfy a common constraint.
For example, let us suppose that the bulk viscosity is given by the following equation,
\begin{equation}
 \zeta_{HH}=\frac{1}{\zeta}\zeta_H^2-\frac{1}{H}(\zeta_H-\alpha \zeta^2-\beta)+\gamma \zeta^3+\frac{\delta}{\zeta}.\end{equation}
Here, $\zeta_{H}=\frac{d\zeta}{dH},$ $ \zeta_{HH}=\frac{d^2\zeta}{dH^{2}}$, $\zeta=\zeta(H)\equiv\zeta(H;\alpha,\beta,\gamma,\delta)$, being $\alpha,\beta,\gamma,\delta$ some constants. This is the well-known Painleve-III (P$_{\text{III}}$) equation and it is important in many physical applications, like for example in soliton theories. Moreover, it is integrable and admits an infinite number of exact solutions. 
In the following, we report some of those specific solutions~\cite{Ablowitz},
\begin{eqnarray}
  \zeta=\zeta(H)&\equiv &\zeta(H;\mu,-\mu\eta^2,\lambda,-\lambda\eta^4)=\eta,\label{66}\\
 \zeta=\zeta(H)&\equiv &\zeta(H;0,-\mu,0,\mu\eta)=\eta H, \label{67}\\ 
  \zeta=\zeta(H)&\equiv &\zeta(H;2\eta+3,-2\eta+1,1,-1)=\frac{H+\eta}{H+\eta+1},\\
  \zeta=\zeta(H)&\equiv& \zeta(H;\mu,0,0,-\mu\eta^3)=\eta \sqrt[3]{H},\\
  \zeta=\zeta(H)&\equiv& \zeta(H;0,-2\eta,0,4\eta\mu-\lambda^2)=H[\eta(\ln H)^2+\lambda\ln H+\mu],\\
\zeta=\zeta(H)&\equiv& \zeta(H;-\sigma^2\lambda,0,\sigma^2(\lambda^2-4\eta\mu),0)=\frac{H^{\sigma-1}}{\eta H^{2\sigma}+\lambda H^{\sigma}+\mu},\\
 \zeta=\zeta(H)&\equiv& \zeta(H;\epsilon_1/2,0,0,0)=-\epsilon_2(\ln{\phi})_{H}\label{72}\,,
 \end{eqnarray}
and so on. Here, $\eta\,,\mu\,,\lambda\,,\sigma\,,\epsilon_{1,2}$ are some constants and 
in the last expression
 \begin{equation}
  \phi(H)=H^{\nu}[C_1J_{\nu}(\xi)+C_2Y_{\nu}(\xi)],
 \end{equation}
 where $C_{1,2}=\text{consts}\,,\nu=\epsilon_1/2\,,\xi=\sqrt{\epsilon_1\epsilon_2}H$ and $J_{\nu}(\xi), Y_{\nu}(\xi)$ are Bessel functions. 
Some comments are in order. First of all, we must put some restrictions on the constant parameters in order to satisfy the requirement $\zeta(H)>0$ (for example from (\ref{66}) it follows $\eta>0$).
At second, we note that the fluids which correspond to the bulk viscosities (\ref{66})--(\ref{72}) have different features despite to the fact that they belong to the same class of integrable models. In some cases we recover the models studied in Section {\bf 3} (we are supposing $\omega$ constant). In the specific, the case (\ref{66}) corresponds to the one considered in \S~\ref{3.1} with $\eta=\zeta_0$; the case (\ref{67}) corresponds to the one considered in \S~\ref{3.2} with $\eta=3\tau$. In general, it is interesting to see that if the bulk viscosity is given by the result of a general constraint (like the one here considered), we may generate an infinite number of models with different behaviours and solutions in FRW space-time derived by the system of equations,
\begin{eqnarray}
\left\{\begin{array}{lll}
  0&=&3H^2-\kappa^2\rho\,,\\ \\
0&=&2\dot {H}+\kappa^2(\rho+p)\,,\\ \\
 0&=& p-\omega\rho+3 H\zeta(H)\,,\\ \\
 0&=&\zeta_{HH}-\frac{1}{\zeta}\zeta_H^2+\frac{1}{H}(\zeta_H-\alpha \zeta^2-\beta)-\gamma \zeta^3-\frac{\delta}{\zeta}\,.
\end{array}\right.
\end{eqnarray} 
This is a closed and integrable system. The integrability is induced by the integrability of the P$_{\text{III}}$-equation, which leads to an infinite number of solutions which provide different FRW universe evolutions related to each others by the Darboux and/or B$\ddot{a}$cklund tranformations. 
\section{Conclusions}

The study of inhomogeneous viscous fluids in Friedmann-Robertson-Walker universe has become important since the discovery of cosmic acceleration and the related dark energy issue.
From one side, the cosmological data do not exclude a more complex form of the dark energy with respect to the simplest one given by the introduction of the Cosmological Constant in the Einstein's equation; then, many dark energy models have a corresponding representation in the fluid-like form. For example, 
the modifications of gravity may be written in such a form and, as a result, one may make use of the framework of 
General Relativity and the analysis of the theory turns out to be simplified.

In this paper, we have considered
different forms of inhomogeneous viscous fluid in flat Friedmann-Robertson-Walker metric,
investigating the possibility to reproduce the current cosmic acceleration with a different future evolution with respect to the Cosmological Constant case. 
We have seen that, like for the case of perfect fluids, it is always possible to obtain fluids for the accelerated universe which violate the strong energy condition: these fluid models may be free of finite-future time singularities or may admitt the presence of them, depending on the values of EoS paramters and of the coefficients of the bulk viscosities. Despite to the fact that in the case of perfect fluids the only singularity which is realized is the Big Rip, by considering more general forms of fluid, other types of singularities may emerge.
We developed a general investigation, by considering the EoS paramter constant and not a constant separately and, inside to such cases, different forms of bulk viscosity. In the last part of our work, we also presented a class of
``integrable'' viscous fluid models whose viscosities obey to a common integrable equation.
In this way, we can generate an infinite number of models with different behaviours and solutions in FRW space-time.

Our approach may be developed as classical analog of landscape in 
analogy with Ref.~\cite{OdOd}.
Other works on inhomogeneous viscous fluids and the dark energy issue have been presented in Refs.~\cite{uno}--\cite{otto} and in Ref.~\cite{LittleRip} for viscous fluids applied to the study of Little Rip cosmology.


\end{document}